\documentclass[prd,twocolumn,floatfix,nofootinbib,amsmath,amssymb,floatfix]{revtex4} 
\usepackage{graphicx,color,dcolumn,booktabs,bm,mathrsfs}
\usepackage{longtable,lscape}
\usepackage{txfonts}
\usepackage{overpic}
\usepackage{amssymb}
\usepackage{indentfirst}
\usepackage{feynmf}   
\usepackage{slashed}  
\usepackage{cases}
\usepackage{color}
\usepackage{multirow}
\usepackage{epstopdf}
\usepackage{CJK}

\usepackage[colorlinks,
            citecolor=blue,
            anchorcolor=red,
            menucolor=red,
            linkcolor=red,
            filecolor=red,
            runcolor=red,
            urlcolor=blue,
            frenchlinks=red]{hyperref}
\usepackage{subfigure}
\usepackage{xcolor}

\newcommand{\tcsn}{T_{c\bar{s}0}(2900)^0}
\newcommand{\tcspp}{T_{c\bar{s}0}(2900)^{++}}
\newcommand{\tcs}{T_{c\bar{s}0}(2900)}
\newcommand{\be}{\begin{equation}} 
\newcommand{\ee}{\end{equation}}
\newcommand{\bea}{\begin{eqnarray}} 
\newcommand{\eea}{\end{eqnarray}}

\begin{document}

\title{Coupled-channel $D^\ast K^\ast -D_s^\ast \rho$ interactions and the origin of $T_{c\bar{s}0}(2900)$}

\author{Man-Yu Duan$^{1}$}\email{duanmy@seu.edu.cn}
\author{Meng-Lin Du$^{2}$}\email{du.ml@uestc.edu.cn}
\author{Zhi-Hui Guo$^{3}$}\email{zhguo@hebtu.edu.cn}
\author{En Wang$^{4}$}\email{wangen@zzu.edu.cn}
\author{Dian-Yong Chen$^{1,5}$}\email{chendy@seu.edu.cn}

\affiliation{$^1$School of Physics, Southeast University, Nanjing 210094, China\\
$^2$School of Physics, University of Electronic Science and Technology of China, Chengdu 611731, China\\
$^3$Department of Physics and Hebei Key Laboratory of Photophysics Research and Application, Hebei Normal University, Shijiazhuang 050024, China\\ 
$^4$School of Physics and Microelectronics, Zhengzhou University, Zhengzhou, Henan 450001, China\\
$^5$Lanzhou Center for Theoretical Physics, Lanzhou University, Lanzhou 730000, P. R. China}

\begin{abstract}

Motivated by the recent observation of $T_{c\bar{s}0}(2900)^0$ and $T_{c\bar{s}0}(2900)^{++}$ in the $D_s \pi$ invariant mass distributions, we investigate $D^{\ast}K^{\ast}$ interactions in a coupled-channel approach. We show that the relativistic corrections could be significant for the energy far away from the threshold. Within the hidden local symmetry formalism, a sizable attraction interaction is found in the $J=0$ isospin triplet sector that can form a bound or a virtual state, which is consistent with the experimentally observed $\tcs$. By reproducing a $D_s^*\rho$-$D^*K^*$ bound/virtual state with the pole mass equal to that of the $\tcs$ measured by LHCb in the sector $(I,J)=(1,0)$, we determine the unknown parameter in the loop function, and then search for possible poles in the sectors of $I=1$, $J=1,$ 2 and $I=0$, $J=0$, 1, 2. The predicted resonances provide a useful reference for the future experimental studies of the $(C,S)=(1,1)$ systems and can be also helpful to unravel the nature of the $\tcs$.

\end{abstract}

\maketitle


\begin{figure*}[thb]
  \centering
  \subfigure{\includegraphics[scale=0.45]{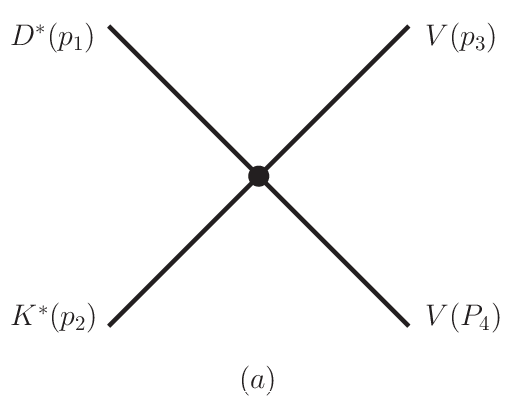}\label{c.eps}}
  \subfigure{\includegraphics[scale=0.45]{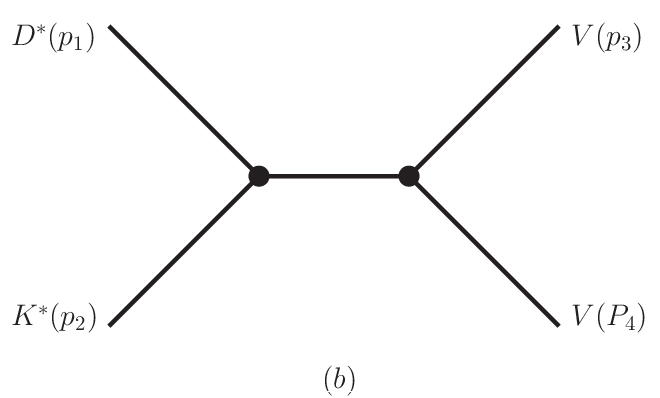}\label{t.eps}}
  \subfigure{\includegraphics[scale=0.45]{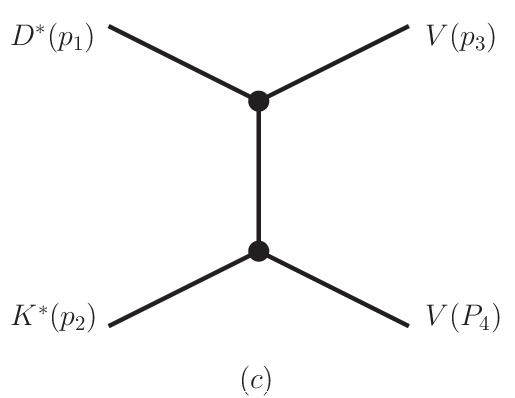}\label{u.eps}}
  \subfigure{\includegraphics[scale=0.45]{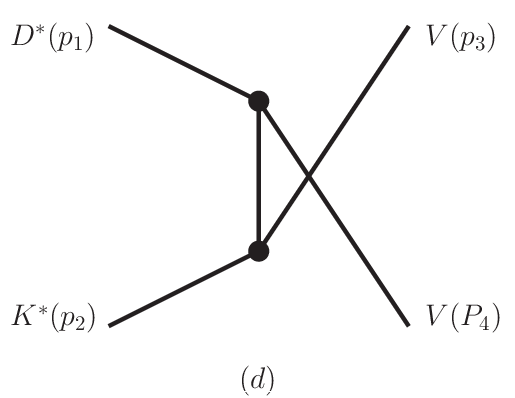}\label{s.eps}}
  \caption{The sketch diagrams for $D^\ast K^\ast \to V V$. Diagrams (a), (b), (c), and (d) correspond to contact interaction, $s$-channel, $t$-channel, and $u$-channel vector meson exchanged interactions, respectively.}
  \label{fig:mechanism}
  \end{figure*}

\section{INTRODUCTION}
\label{sec:INTRODUCTION}

In 2022, the LHCb Collaboration reported two new states, $T_{c\bar{s}0}(2900)^0$ and $T_{c\bar{s}0}(2900)^{++}$, in the decays $B^0 \to \bar{D}^0 D_s^+ \pi^-$ and $B^+ \to D^- D_s^+ \pi^+$, respectively~\cite{LHCb:2022xob,LHCb:2022bkt}. The two states decay to $D_s^+\pi^-$ and $D_s^+\pi^+$, respectively, which implies that their minimal quark contents are $[c\bar{s}\bar{u}d]$ and $[c\bar{s}u\bar{d}]$. Both states are found to have spin-parity $J^P=0^+$ and their resonance parameters extracted from the relativistic Breit-Wigner fits by LHCb are~\cite{LHCb:2022xob,LHCb:2022bkt},
\begin{eqnarray}
m_{T_{c\bar{s}0}(2900)^{0}}&=&(2892\pm14\pm15)~\mathrm{MeV}\ ,\nonumber\\
\Gamma_{T_{c\bar{s}0}(2900)^{0}}&=&(119\pm26\pm13)~\mathrm{MeV}\ ,\nonumber\\
m_{T_{c\bar{s}0}(2900)^{++}}&=&(2921\pm17\pm20)~\mathrm{MeV}\ ,\nonumber\\
\Gamma_{T_{c\bar{s}0}(2900)^{++}}&=&(137\pm32\pm17)~\mathrm{MeV}\ ,
\end{eqnarray}
which are compatible with each other within uncertainties. By assuming the two resonances belong to the same isospin triplet, the common mass and width of $T_{c\bar{s}0}(2900)$ are fitted to be~\cite{LHCb:2022xob,LHCb:2022bkt},
\begin{eqnarray}
m_{T_{c\bar{s}0}(2900)}&=&(2908\pm11\pm20)~\mathrm{MeV}\ ,\nonumber\\
\Gamma_{T_{c\bar{s}0}(2900)}&=&(136\pm23\pm13)~\mathrm{MeV}\ .
\label{respara}
\end{eqnarray}

The discovery of $\tcspp$ and $\tcsn$ quickly spurred a number of theoretical studies as the former state is the first observation of a doubly charged open-charm tetraquark state. Unraveling their origin is important to understanding the strong interaction. The proximity of the $D^*K^*$ and the $D_s^*\rho$ thresholds to the mass of $T_{c\bar{s}0}(2900)$ suggests that these two-hadron channels could play important roles in the dynamics of the $\tcs$ states, hinting to a hadronic molecular interpretation of the two states \cite{Agaev:2022eyk, Yue:2022mnf, Chen:2022svh, Duan:2023qsg,Lyu:2023jos}. The alternative interpretation as compact tetraquark states with quark contents $[c\bar{s}\bar{u}d]$ and $[c\bar{s}u\bar{d}]$ is studied in Refs.~\cite{Yang:2023evp, Lian:2023cgs, Jiang:2023rcn, Liu:2022hbk, Dmitrasinovic:2023eei, Ortega:2023azl}. In addition to being a genuine state, the $\tcs$ structure is also proposed to be merely a threshold cusp effect from the interaction between the $D^*K^*$ and $D_s^*\rho$ channels \cite{Molina:2022jcd} or the kinetic effect from a triangle singularity~\cite{Ge:2022dsp}.

In Ref.~\cite{Molina:2022jcd}, the $D^*K^*$ (and $D^*\bar{K}^*$) system was investigated within the framework of the extended hidden local symmetry approach to SU(4) to incorporate charmed mesons. In that work, the nonrelativistic approximation, i.e. $\vec{p}/M_V\to 0$ with $\vec{p}$ the three-momentum of the involved states~\cite{Molina:2008jw}, was taken. It was found that, in the isovector sector with $(C,S)=(1,1)$, being $C$ and $S$ the charmness and strangeness numbers in order, i.e. the $D^*K^*$-$D_s^*\rho$ coupled-system, while a bound state can be found for $J=2$ sector, only cusp effects are observed for $J=0$ and 1 sectors though attractive potentials appear in these two cases~\cite{Molina:2022jcd}. Due to the relatively strong attractive potentials, three deep bound states are found for $J=0,1,$ and 2 in the isoscalar sector of $(C,S)=(1,1)$. In this sector also sits the well-known $D_{s0}^*(2317)$ discovered in the inclusive $D_s^+\pi^0$ invariant mass distribution from $e^+e^-$ annihilation data by the BaBar Collaboration in 2003~\cite{BaBar:2003oey}. The $D_{s0}^*(2317)$ is suggested as dominantly a $DK$ hadronic molecule \cite{Barnes:2003dj, Chen:2004dy, Guo:2017jvc, Cleven:2010aw, Liu:2012zya, Du:2017ttu, Yang:2021tvc,Liu:2022dmm} due to that it is located far below the conventional quark model expectation \cite{Godfrey:2003kg} and just below the $DK$ threshold. The heavy-quark symmetry implies that the $D^*K$ interaction is identical to the $DK$ interaction up to $\mathcal{O}(\Lambda_\text{QCD}/m_c)$ and thus there should exsit a $D^*K$ molecule which is identified to the $D_{s1}(2460)$ observed in the $D_s^{*+}\pi^0$ mass distribution \cite{CLEO:2003ggt}. The molecular interpretations of the $D_{s0}^*(2317)$ and $D_{s1}(2460)$ as $DK$ and $D^*K$ molecules are supported by the observation that $M_{D_{s1}(2460)}-M_{D^*} \simeq M_{D_{s0}^*(2317)}-M_D$. Though both the $D^{(*)}K$ and the $D^*K^*$ systems are attractive and  generate poles in the isoscalar sector~\cite{Molina:2022jcd}, it is worth stressing that their origins are different. The $D_{s0}^*(2317)$ and $D_{s1}(2460)$ emerge as a consequence of the spontaneous chiral symmetry breaking of QCD, which constrains the $D^{(*)}K$ interaction since the $K$ is the corresponding Goldstone boson. However, the deep bound states found in the $I=0$ $D^*K^*$ systems are obtained with the potentials derived from the hidden gauge formalism  in the nonrelativistic approximation \cite{Molina:2022jcd}. 

It is remarked that for the two-body scattering processes with pure light-flavor vectors, such as $\rho\rho\to\rho\rho$, the deep bound states generated with the nonrelativistic approximation in Ref.~\cite{Molina:2008jw,Geng:2008gx} is untenable due to the neglect of the relativistic effects~\cite{Gulmez:2016scm,Du:2018gyn}. The same issue could also exist in the $D^*_{(s)}V$ (with $V$ the light-flavor vector) scattering. Therefore one of the key motivations of this work is to  study to which level the $D^*_s\rho$-$D^*K^*$ system could be affected by including relativistic effects. For energy regions far away from the two-body threshold, the nonrelativistic approximation is questioned as the relativistic corrections could be significant. Therefore, a relativistically covariant formalism is employed in Refs.~\cite{Gulmez:2016scm,Du:2018gyn}, which leads to an ``unphysical'' left-hand cut with the on-shell factorization. It is worth emphasizing that this issue is caused by the on-shell factorization and can be overcome by the Lippmann-Schwinger equation or (the first iterated solution of) the $N/D$ dispersion relation \cite{Gulmez:2016scm,Du:2018gyn}. It is found that, while the poles generated in the very vicinity of the threshold are consistent between the relativistic and nonrelativistic formalism, those found far away from the threshold in the nonrelativistic approximation are unreliable.  Moreover, the corrections from the higher-order effective Lagrangians to the derived potentials at the energy regions far away from the threshold could be sizable. Hence, as a conservative estimate, in this work, we will try to reinvestigate the $D^*K^*$ interactions with a relativistically covariant formalism and restrict ourselves to the energy region above the corresponding left-hand cuts developed by the vector-exchanging diagrams. 
 
The quantum numbers of the $\tcs$ are determined to be $I=1$ and $J^P=0^+$ \cite{LHCb:2022xob,LHCb:2022bkt}. As indicated in Ref.~\cite{Molina:2010tx}, in the isovector sector of the $(C,S)=(1,1)$, the potential for $D_s^*\rho \to D_s^*\rho$ vanishes and that for $D^*K^*\to D^*K^*$ is neglegible. A sizable potential for $D^*K^*\to D_s^*\rho$ leads to an attractive effect in this coupled-channel. It is easy to see the conclusion from a  combination of the two-body states, i.e. $|\Psi\rangle_\pm \simeq \frac{1}{\sqrt{2}} | D^*K^*\rangle \pm |D_s^*\rho\rangle$. In particular, a negative potential for $D^*K^*\to D_s^*\rho$ suggests that while the potential for $|\Psi\rangle_+$ to $|\Psi\rangle_+$ is attractive, that for $|\Psi\rangle_-$ to $|\Psi\rangle_-$ is repulsive. The transition between $|\Psi\rangle_+$ and $|\Psi\rangle_-$ vanishes, which corresponds to diagonalization of the potentials matrix. The attraction could generate a bound state if the strength is sufficiently strong.  In particular, at the $D^*K^*$ threshold, the potential for $D^*K^*\to D_s^*\rho$ is $-6.8g^2$, where $g=M_\rho/2f_\pi=4.17$ with $f_\pi=93$~MeV. It is easy to see that the attraction effect is sizable. However, to determine whether a bound state can be formed, an estimate of the subtraction constant, $\alpha(\mu)$, is required for the two-point loop function evaluated using dimensional regularization. In Ref.~\cite{Molina:2010tx}, by setting the renormalization scale $\mu=1500$ MeV and $\alpha=-1.6$, no pole was obtained for $J=0$, but only a cusp was observed in the $D_s^*\rho$ threshold. It is worth noticing that the cusp in the threshold may indicate a relatively strong interaction, and whether a bound state can be formed is sensitive to the choice of the subtraction $\alpha$~\cite{Duan:2022upr,Duan:2021pll}. 
We will see below that a bound state can be found by slightly changing the value of $\alpha$. For instance, by choosing $\alpha=-1.65$, a pole can be found at around 2886~MeV in the physical Riemann sheet (RS), which is identified as a bound state. As a matter of fact, with $\alpha=-1.6$, a pole below the $D_s^*\rho$ threshold at around {2886}~MeV is found in the unphysical RS, which corresponds to a virtual state and shows up as a cusp at the threshold in the amplitude of the physical RS. Without prior knowledge of the subtraction $\alpha(\mu)$ (although its natural size is discussed in Ref.~\cite{Oller:2000fj}), the loop function can be estimated by the hard-cutoff regularization with a natural value of the cutoff $q_\text{max}\sim M_V$. The value of $\alpha(\mu)$ then can be estimated by matching the loop functions evaluated with the two methods at a certain point, e.g., the threshold. We will show that in a reasonable range of the cutoff $q_\text{max}$, the determined $\alpha$ could lead to a bound state or a virtual state. On the other hand, the coupling $g=M_\rho/2f_\pi=4.17$ is used in Ref.~\cite{Molina:2010tx}. When an SU(3) average mass of the vectors is employed, i.e. $g=4.60$ \cite{Du:2018gyn}, a bound state at around 2873 MeV can be found even with $\alpha =-1.6$. As a consequence, the observed $\tcs$ is consistent with a $D^*K^*$-$D_s^*\rho$ bound state/virtual state with $J=0$. In this work, we will take  advantage of the LHCb measurement to determine the $\alpha$ by assuming the $\tcs$ as a $D^*K^*$-$D_s^*\rho$ bound state/virtual state with its pole mass equal to the value of Eq.~\eqref{respara}. Since we only focus on the origin of possible dynamically generated states of $D^*K^*$ interactions, we will not consider its width due to the transitions to inelastic two-body channels, e.g. $DK$ and $D_s\pi$, three-body and four-body channels, which are supposed to affect its pole mass insignificantly~\cite{Molina:2008jw,Gulmez:2016scm}.

This paper is organized as follows. In Section~\ref{sec:FORMALISM} we derive the relativistically covariant partial-wave potentials, and demonstrate the formalism to calculate the unitarized scattering amplitudes. The numerical results and discussions are presented in Section~\ref{sec:RESULTS}. Section~\ref{sec:SUMMARY} is devoted to a short summary.


\renewcommand\arraystretch{1.5}
\begin{table*}[htb]
   \begin{center}
  \caption{The coefficients of amplitudes for the contact term and vector-exchanging term in Eq.~(\ref{eq:totalamp}).}
  \label{tab:coff}
  \setlength{\tabcolsep}{1.9mm}{
  \begin{tabular}{cccccccccccccccccccccccc}
 \toprule[1 pt]
  Isospin  & Channel  & $C_1$  & $C_2$  & $C_s^{\rho}$  & $C_s^{\omega}$  & $C_s^{K^*}$ & $C_s^{D^*}$ & $C_s^{D_s^*}$  & $C_t^{\rho}$  & $C_t^{\omega}$  & $C_t^{K^*}$ & $C_t^{D^*}$ & $C_t^{D_s^*}$ & $C_u^{\rho}$  & $C_u^{\omega}$  & $C_u^{K^*}$ & $C_u^{D^*}$ & $C_u^{D_s^*}$\\
 \midrule[1 pt]
  I = 0   &$D^{\ast}K^{\ast} \to D^{\ast}K^{\ast}$    ~ & 1 & 0 ~ & 0 & 0 & 0 & 0 & 2 ~ & $\frac{3}{2}$ & $\frac{1}{2}$ & 0  & 0 & 0 ~ & 0  & 0 & 0 & 0 & 0  \\
  \,   &$D^{\ast}K^{\ast} \to D_s^{\ast} \omega$   ~ & 0 & -$\frac{1}{2}$ ~ & 0 & 0 & 0 & 0 & 0 ~ & 0 & 0  & -1  & 0 & 0 ~ & 0 & 0 & 0  & -1 & 0  \\
  \,   &$D^{\ast}K^{\ast} \to D_s^{\ast} \phi$       ~ & $\frac{\sqrt{2}}{2}$ & 0 ~ & 0 & 0 & 0 & 0 & $\sqrt{2}$ ~ & 0 & 0 & $\sqrt{2}$ & 0 & 0 ~ & 0 & 0 & 0  & 0 & 0  \\
  \,   &$D_s^{\ast} \phi \to D_s^{\ast} \phi$       ~ & $\frac{1}{2}$ & -$\frac{1}{2}$ ~ & 0 & 0 & 0 & 0 & 1 ~ & 0 & 0 & 0 & 0 & 0 ~ & 0 & 0  & 0 & 0 & -1  \\
  I = 1   &$D^{\ast}K^{\ast} \to D^{\ast}K^{\ast}$   ~ & 0 & 0 ~ & 0 & 0 & 0 & 0 & 0 ~ & -$\frac{1}{2}$ & $\frac{1}{2}$ & 0 & 0 & 0 ~ & 0 & 0 & 0 & 0 & 0  \\
  \,   &$D^{\ast}K^{\ast} \to D_s^{\ast} \rho$      ~ & 0 & $\frac{1}{2}$ ~ & 0 & 0 & 0 & 0 & 0 ~ & 0 & 0 & 1 & 0 & 0 ~ & 0 &0 & 0 & 1 & 0  \\
 \bottomrule[1 pt]
  \end{tabular}}
  \end{center}
  \end{table*}

 \begin{figure*}[htb]
 \begin{tabular}{rrr}
 \includegraphics[scale=0.7]{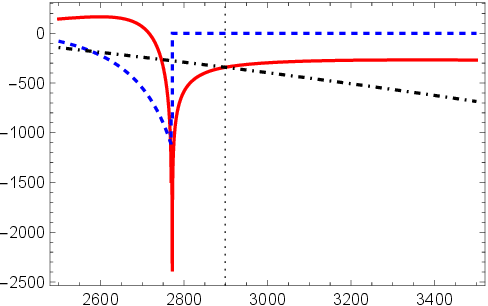} \put(-15,20){\scriptsize{(a1)}} \put(-37,10){\scriptsize{$I=0$, $J=0$}} \put(-87,10){\scriptsize{\textcolor{gray}{$D^* K^*$}}} \put(-176,35){\normalsize\rotatebox{90}{$V_{D^*K^* \to D^*K^*}$}} &
 \includegraphics[scale=0.7]{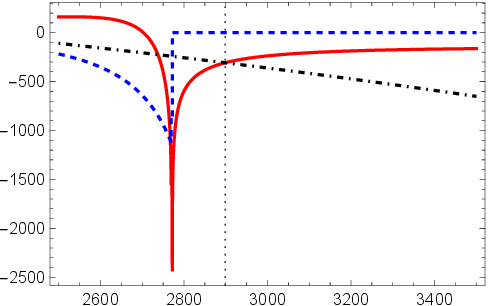} \put(-15,20){\scriptsize{(a2)}} \put(-37,10){\scriptsize{$I=0$, $J=1$}} \put(-87,10){\scriptsize{\textcolor{gray}{$D^* K^*$}}} &
 \includegraphics[scale=0.7]{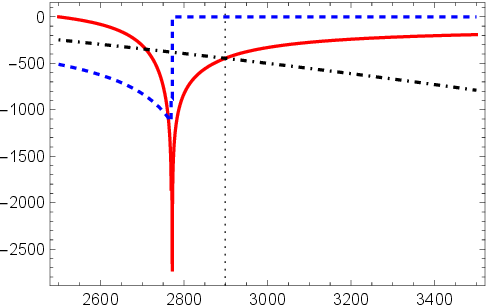} \put(-15,20){\scriptsize{(a3)}} \put(-37,10){\scriptsize{$I=0$, $J=2$}} \put(-87,10){\scriptsize{\textcolor{gray}{$D^* K^*$}}}
 \\
 \includegraphics[scale=0.7]{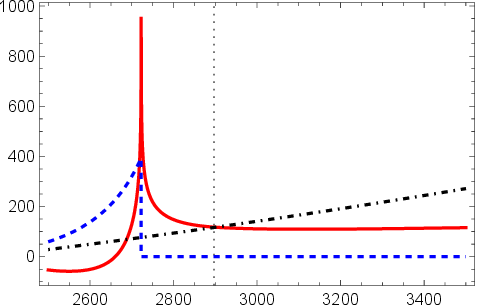} \put(-15,85){\scriptsize{(b1)}} \put(-37,95){\scriptsize{$I=0$, $J=0$}} \put(-103,95){\scriptsize{\textcolor{gray}{$D_s^*\omega$}}} \put(-87,95){\scriptsize{\textcolor{gray}{$D^* K^*$}}} \put(-176,35){\normalsize\rotatebox{90}{$V_{D^*K^* \to  D^*_s \omega}$}} &
 \includegraphics[scale=0.7]{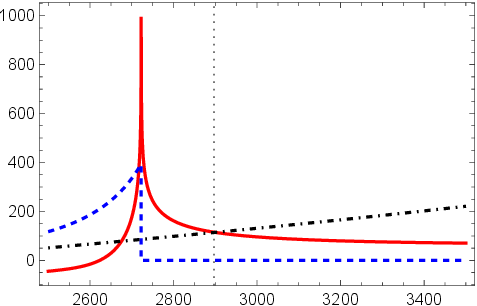} \put(-15,85){\scriptsize{(b2)}} \put(-37,95){\scriptsize{$I=0$, $J=1$}} \put(-103,95){\scriptsize{\textcolor{gray}{$D_s^*\omega$}}} \put(-87,95){\scriptsize{\textcolor{gray}{$D^* K^*$}}} &
 \includegraphics[scale=0.7]{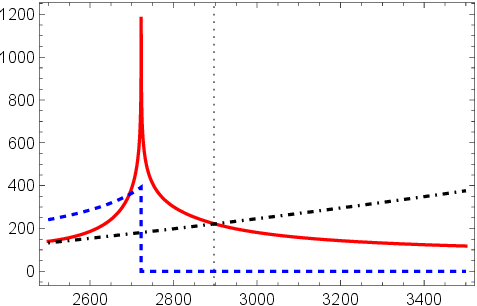} \put(-15,85){\scriptsize{(b3)}} \put(-37,95){\scriptsize{$I=0$, $J=2$}} \put(-103,95){\scriptsize{\textcolor{gray}{$D_s^*\omega$}}} \put(-87,95){\scriptsize{\textcolor{gray}{$D^* K^*$}}}
  \\
  \includegraphics[scale=0.7]{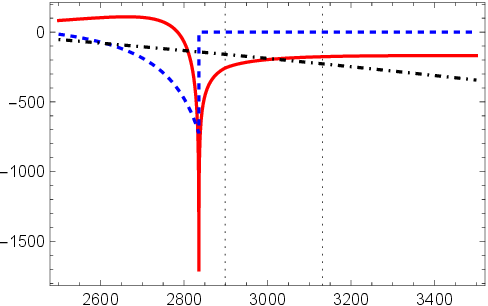} \put(-144,20){\scriptsize{(d1)}} \put(-145,10){\scriptsize{$I=0$, $J=0$}} \put(-54,11){\scriptsize{\textcolor{gray}{$D_s^*\phi$}}} \put(-87,10){\scriptsize{\textcolor{gray}{$D^* K^*$}}} \put(-176,35){\normalsize\rotatebox{90}{$V_{D^*K^* \to D^*_s \phi}$}} &
  \includegraphics[scale=0.7]{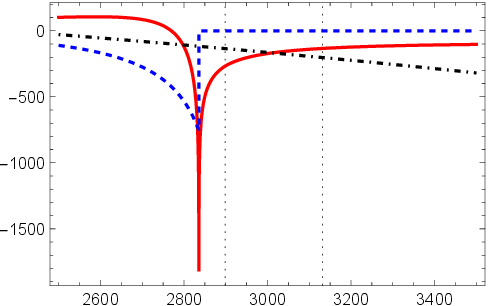} \put(-144,20){\scriptsize{(d2)}} \put(-145,10){\scriptsize{$I=0$, $J=1$}} \put(-54,11){\scriptsize{\textcolor{gray}{$D_s^*\phi$}}} \put(-87,10){\scriptsize{\textcolor{gray}{$D^* K^*$}}} &
  \includegraphics[scale=0.7]{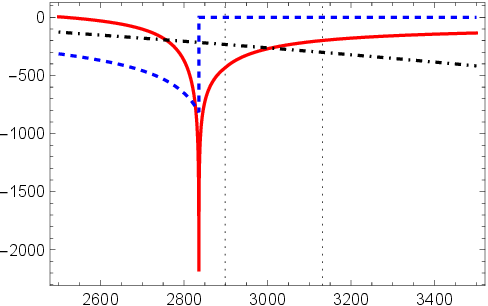} \put(-144,20){\scriptsize{(d3)}} \put(-145,10){\scriptsize{$I=0$, $J=2$}} \put(-54,11){\scriptsize{\textcolor{gray}{$D_s^*\phi$}}} \put(-87,10){\scriptsize{\textcolor{gray}{$D^* K^*$}}}
   \\
  \includegraphics[scale=0.7]{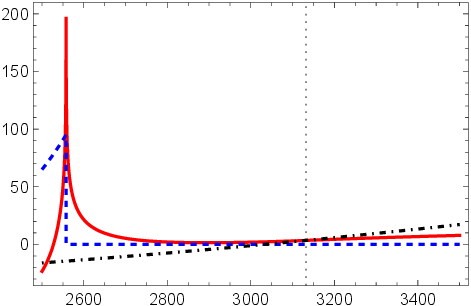} \put(-15,85){\scriptsize{(e1)}} \put(-37,95){\scriptsize{$I=0$, $J=0$}} \put(-69,95){\scriptsize{\textcolor{gray}{$D_s^*\phi$}}} \put(-176,35){\normalsize\rotatebox{90}{$V_{D^*_s \phi \to D^*_s \phi}$}} \put(-100,-15){\normalsize{$\sqrt{s}$ (MeV)}} &
  \includegraphics[scale=0.7]{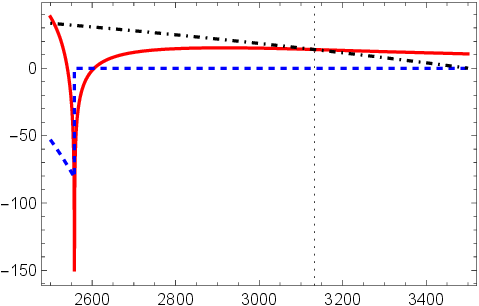} \put(-15,20){\scriptsize{(e2)}} \put(-37,10){\scriptsize{$I=0$, $J=1$}} \put(-69,11){\scriptsize{\textcolor{gray}{$D_s^*\phi$}}} \put(-100,-15){\normalsize{$\sqrt{s}$ (MeV)}} &
  \includegraphics[scale=0.7]{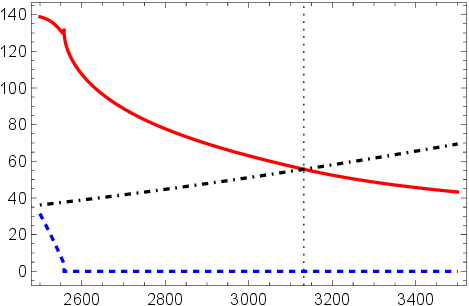} \put(-15,85){\scriptsize{(e3)}} \put(-37,95){\scriptsize{$I=0$, $J=2$}} \put(-69,95){\scriptsize{\textcolor{gray}{$D_s^*\phi$}}} \put(-100,-15){\normalsize{$\sqrt{s}$ (MeV)}}
   \\
  \end{tabular}
  \caption{$S$-wave potentials $V^{(IJ)}$ defined by Eq.~\eqref{eq:pwd} for $VV \to VV$ with $I=0$.  The red solid and blue dashed curves are the real and imaginary parts of the potentials, respectively, while the black dot-dashed curves are the nonrelativistic potentials in Ref.~\cite{Molina:2010tx}. The black dotted lines correspond to the $VV$ thresholds.}
  \label{fig:V0}
  \end{figure*}

  \begin{figure*}[htb]
  \begin{tabular}{rrr}
   \includegraphics[scale=0.7]{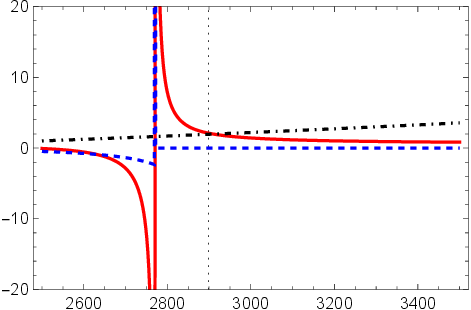} \put(-15,20){\scriptsize{(f1)}} \put(-37,10){\scriptsize{$I=1$, $J=0$}} \put(-87,10){\scriptsize{\textcolor{gray}{$D^* K^*$}}}  \put(-176,35){\normalsize\rotatebox{90}{$V_{D^*K^* \to D^*K^*}$}} &
   \includegraphics[scale=0.7]{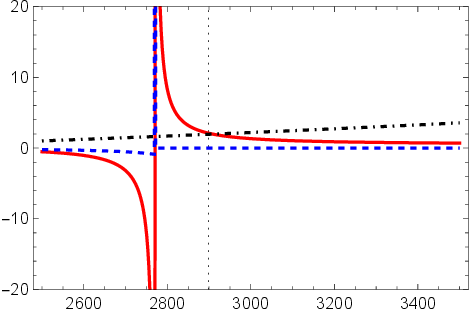}\put(-15,20){\scriptsize{(f2)}} \put(-37,10){\scriptsize{$I=1$, $J=1$}} \put(-87,10){\scriptsize{\textcolor{gray}{$D^* K^*$}}} &
   \includegraphics[scale=0.7]{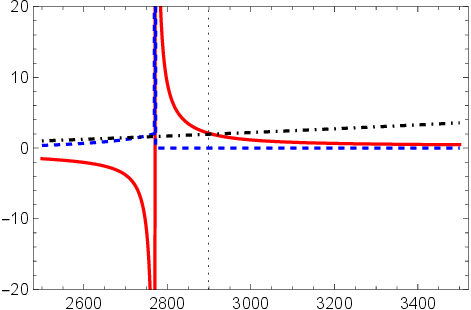}\put(-15,20){\scriptsize{(f3)}} \put(-37,10){\scriptsize{$I=1$, $J=2$}} \put(-87,10){\scriptsize{\textcolor{gray}{$D^* K^*$}}}
    \\
   \includegraphics[scale=0.7]{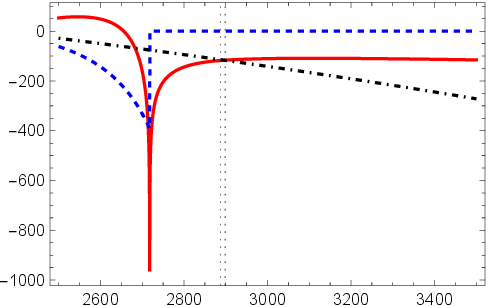} \put(-15,20){\scriptsize{(h1)}} \put(-37,10){\scriptsize{$I=1$, $J=0$}}\put(-87,10){\scriptsize{\textcolor{gray}{$D^* K^*$}}}  \put(-104,11){\scriptsize{\textcolor{gray}{$D_s^* \rho$}}}  \put(-176,35){\normalsize\rotatebox{90}{$V_{D^*K^* \to D^*_s \rho}$}}  \put(-100,-15){\normalsize{$\sqrt{s}$ (MeV)}} &
   \includegraphics[scale=0.7]{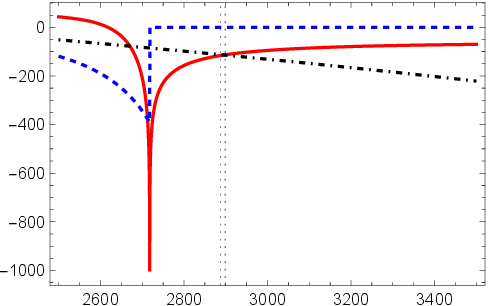} \put(-15,20){\scriptsize{(h2)}} \put(-37,10){\scriptsize{$I=1$, $J=1$}}\put(-87,10){\scriptsize{\textcolor{gray}{$D^* K^*$}}}  \put(-104,11){\scriptsize{\textcolor{gray}{$D_s^* \rho$}}}  \put(-100,-15){\normalsize{$\sqrt{s}$ (MeV)}} &
   \includegraphics[scale=0.7]{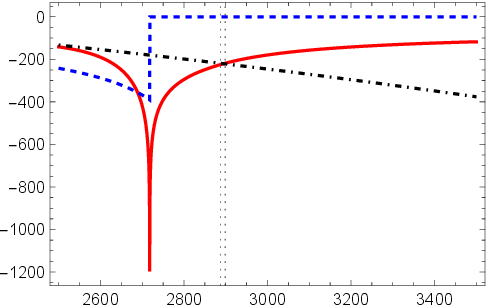} \put(-15,20){\scriptsize{(h3)}} \put(-37,10){\scriptsize{$I=1$, $J=2$}}\put(-87,10){\scriptsize{\textcolor{gray}{$D^* K^*$}}}  \put(-104,11){\scriptsize{\textcolor{gray}{$D_s^* \rho$}}}  \put(-100,-15){\normalsize{$\sqrt{s}$ (MeV)}}
    \\
   \end{tabular}
   \caption{The same as in Fig.~\ref{fig:V0}, but for $I=1$. }  
   \label{fig:V1}
  \end{figure*}

\section{FORMALISM}
\label{sec:FORMALISM}

In order to make a close comparison with the nonrelativistic treatment of the $D^*K^*$ system as done in Ref.~\cite{Molina:2010tx}, it is convenient to take the same theoretical model employed in the former reference, i.e. a straightforward extension of the hidden local symmetry formalism to SU(4) to include charmed vectors, to investigate the $D^*K^*$ interactions, although this model could be somewhat oversimplified to deal with the interaction vertices of the charmed and light-flavor mesons. Nevertheless, as stressed in Section~\ref{sec:INTRODUCTION}, one aim here is to study the relativistic effects in the $D_s^*\rho$-$D^*K^*$ system in a covariant manner within the coupled-channel approach. 
The Lagrangian describing the interactions among vector mesons reads~\cite{Bando:1984ej,Bando:1987br,Molina:2010tx},
\begin{equation}
\mathcal{L}=-\frac{1}{4}\langle V_{\mu\nu}V^{\mu\nu}\rangle\ ,
\label{eq:lvv}
\end{equation}
where the symbol $\langle \dots \rangle$ stands for the trace over SU(4) flavor space, and the tensor $V_{\mu\nu}$ is defined as
\begin{equation}
V_{\mu\nu}=\partial_{\mu}V_{\nu}-\partial_{\nu}V_{\mu}-ig[V_{\mu},V_{\nu}]\ ,
\label{eq:vectensor}
\end{equation}
with the coupling constant $g=4.17$ as in Ref.~\cite{Molina:2010tx}. The vector meson matrix $V_{\mu}$ is
\begin{equation}
V_\mu=\left(
\begin{array}{cccc}
\frac{\omega}{\sqrt{2}}+\frac{\rho^0}{\sqrt{2}} & \rho^+ & K^{\ast+}&\bar{D}^{\ast0}\\
\rho^- &\frac{\omega}{\sqrt{2}}-\frac{\rho^0}{\sqrt{2}} & K^{\ast0}&D^{\ast-}\\
K^{\ast-} & \bar{K}^{\ast0} &\phi&D^{\ast-}_s\\
D^{\ast0}&D^{\ast+}&D^{\ast+}_s&J/\psi
\end{array}
\right)_\mu\ .
\label{eq:vfields}
\end{equation}
By expanding the effective Lagrangian in Eq. (\ref{eq:lvv}), one obtains two types of vector interaction vertices, which are the four-vector contact term [Fig.~\ref{fig:mechanism}-(a)] and the three-vector vertices responsible for the vector-exchange interactions [Fig.~\ref{fig:mechanism}-(b-d)], respectively. 

As for the four-vector contact interaction, the corresponding Lagrangian is 
\begin{equation}
\mathcal{L}^{(c)}=\frac{g^2}{2}\langle V_{\mu}V_{\nu}V^{\mu}V^{\nu}-V_{\nu}V_{\mu}V^{\mu}V^{\nu}\rangle\ .
\label{eq:contact}
\end{equation}
One can obtain the corresponding amplitude, which is given by,
\bea
\mathcal{A}^{(c)} = C_1 \mathcal{A}^{(c)}_1 + C_2 \mathcal{A}^{(c)}_2 ,
\eea
with channel-dependent coefficients $C_1$ and $C_2$, and 
\begin{eqnarray}
\mathcal{A}^{(c)}_1&=&2g^2 (\epsilon_1 \cdot \epsilon_2 ~\epsilon_3^{\ast} \cdot \epsilon_4^{\ast}+\epsilon_1 \cdot \epsilon_3^{\ast} ~\epsilon_2 \cdot \epsilon_4^{\ast}-2\epsilon_1 \cdot  \epsilon_4^{\ast} ~\epsilon_2 \cdot \epsilon_3^{\ast}  )\ ,\nonumber\\
\mathcal{A}^{(c)}_2&=&2g^2 (\epsilon_1 \cdot \epsilon_3^{\ast} ~\epsilon_2 \cdot \epsilon_4^{\ast}+\epsilon_1 \cdot  \epsilon_4^{\ast} ~\epsilon_2 \cdot \epsilon_3^{\ast}-2\epsilon_1 \cdot \epsilon_2 ~\epsilon_3^{\ast} \cdot \epsilon_4^{\ast}  )\ , \nonumber\\
\label{eq:ampc}
\end{eqnarray}
where the indices 1, 2, 3, and 4 correspond to the particles with the momenta $p_1$, $p_2$, $p_3$, and $p_4$ in Fig.~\ref{fig:mechanism}-(a), respectively, the $\epsilon_i^{(*)}$ is the polarization vector of the $i$th particle\footnote{The concrete expression of polarization vector can be found in the Appendix A of Ref.~\cite{Gulmez:2016scm}.}, the dot indicates the scalar product, and the superscript $(c)$ stands for the contact term.

The vector-exchange diagrams are described by the Lagrangian
  \begin{eqnarray}
{\cal L}^{(3V)}=ig\langle V^\nu\partial_\mu V_\nu V^\mu-\partial_\nu V_\mu
V^\mu V^\nu\rangle \nonumber\\
=ig\langle (V^\mu\partial_\nu V_\mu -\partial_\nu V_\mu
V^\mu) V^\nu\rangle\ .
\label{l3Vsimp}
\end{eqnarray}
The $t$-channel amplitude exchanging the vector $V_{ex}$ with mass $M_{ex}$ corresponding to Fig.~\ref{fig:mechanism}-(c) has the form
\begin{eqnarray}
C_t^{V_{ex}}{\cal A}^{(t)}_{V_{ex}}&=&C_t^{V_{ex}}\frac{g^2}{t-M_{ex}^2}\Bigg[ \epsilon_1 \cdot \epsilon_3^{\ast} ~\epsilon_2 \cdot \epsilon_4^{\ast} \nonumber\\
& &\times~\bigg(s-u+\frac{(M_1^2-M_3^2)(M_2^2-M_4^2)}{M_{ex}^2}\bigg) \nonumber\\
& &-~4~ \epsilon_1 \cdot \epsilon_3^{\ast} ~(p_1 \cdot \epsilon_2 ~p_2 \cdot \epsilon_4^{\ast}+p_1 \cdot \epsilon_4^{\ast} ~p_4 \cdot \epsilon_2) \nonumber\\
& &+~4~ p_1 \cdot \epsilon_3^{\ast} ~(\epsilon_1 \cdot \epsilon_2 ~p_2 \cdot \epsilon_4^{\ast}+\epsilon_1 \cdot \epsilon_4^{\ast} ~p_4 \cdot \epsilon_2) \nonumber\\
& &-~4~ \epsilon_2 \cdot \epsilon_4^{\ast} ~(p_1 \cdot \epsilon_3^{\ast} ~p_2 \cdot \epsilon_1+p_2 \cdot \epsilon_3^{\ast} ~p_3 \cdot \epsilon_1) \nonumber\\
& &+~4~ p_3 \cdot \epsilon_1 ~(\epsilon_2 \cdot \epsilon_3^{\ast} ~p_2 \cdot \epsilon_4^{\ast}+\epsilon_3^{\ast} \cdot \epsilon_4^{\ast} ~p_4 \cdot \epsilon_2) \Bigg]~,  \nonumber\\
\label{ampt}
\end{eqnarray}
where $C^{V_{ex}}_t$ is a channel-dependent coefficient, $M_1$, $M_2$, $M_3$, and $M_4$ stand for the masses of the particles with the momenta $p_1$, $p_2$, $p_3$, and $p_4$ in Fig.~\ref{fig:mechanism}-(c), respectively, and the Mandelstam variables are defined as $s=(p_1+p_2)^2$, $t=(p_1-p_3)^2$ and $u=(p_1-p_4)^2$, which satisfy the constraint $s+t+u=\sum_{i=1,4} M_i^2$. The $u$-channel exchanging amplitude ${\cal A}^{(u)}_{V_{ex}}$ can be obtained from the expression of ${\cal A}^{(t)}_{V_{ex}}$ by exchanging $p_3 \leftrightarrow p_4$ and $\epsilon_3^{\ast} \leftrightarrow \epsilon_4^{\ast}$. Similarly, the $s$-channel exchanging amplitude ${\cal A}^{(s)}_{V_{ex}}$ can also be obtained from the expression of ${\cal A}^{(t)}_{V_{ex}}$ by performing the exchange $p_2 \leftrightarrow -p_3$ and $\epsilon_2 \leftrightarrow \epsilon_3^{\ast}$. Then the tree-level scattering amplitude for a certain process is given by
\begin{eqnarray}
{\cal A}&=&C_{1}~{\cal A}^{(c)}_{1}+C_{2}~{\cal A}^{(c)}_{2} \nonumber\\
& &+C_{s}^{V_{ex}}~ {\cal A}^{(s)}_{V_{ex}}+C_{t}^{V_{ex}}~ {\cal A}^{(t)}_{V_{ex}}+C_{u}^{V_{ex}}~ {\cal A}^{(u)}_{V_{ex}}
\label{eq:totalamp}\ ,
\end{eqnarray}
where the $V_{ex}$ runs over all possible exchanging vectors.

In the present work, we focus on the channels with charmness $C=1$ and strangeness $S=1$. In the isoscalar sector, i.e. $I = 0$, three channels, namely $D^{\ast}K^{\ast}$, $D_s^{\ast} \omega$ and $D_s^{\ast} \phi$, are involved. In the isovector sector, we take two channels into account, $D^{\ast}K^{\ast}$ and $D_s^{\ast} \rho$. The tree-level amplitudes for the transitions among those channels with certain isospin are given by Eq.~\eqref{eq:totalamp} with the coefficients collected in Table~\ref{tab:coff}. The transitions of $D_s^{\ast} \omega \to D_s^{\ast} \omega$, $D_s^{\ast} \omega \to D_s^{\ast} \phi$, and $D_s^{\ast} \rho \to D_s^{\ast} \rho$ vanish in the present model, thus are not shown.

By means of the above amplitudes with definite isospin, one can calculate the partial-wave amplitudes 
in the $I J \ell S$ basis (states with definite isospin $I$, total angular momentum $J$, orbital angular momentum $\ell$ and total spin $S$), denoted as $V^{(IJ)}_{\ell S;\bar{\ell}\bar{S}}(s)$ for the transition 
$(I J \bar{\ell}\bar{S})\to (I J \ell S)$. The expression for partial-wave decomposition reads \cite{Gulmez:2016scm}
\begin{eqnarray}
\label{eq:pwd}
V^{(IJ)}_{\ell S;\bar{\ell}\bar{S}}(s)&=&\frac{Y_{\bar{\ell}}^{0}(\hat{\mathbf{z}})}{2J+1}\sum_{\scriptsize{\sigma_1,\sigma_2,\bar{\sigma}_1,\bar{\sigma}_2,m}} \int d\hat{\mathbf{p}}'' Y_{\ell}^m(\mathbf{p}'')^* (\sigma_1 \sigma_2 M | s_1 s_2 S)  \nonumber\\
& &\times~(m M \bar{M}|\ell S J)(\bar{\sigma}_1\bar{\sigma}_2\bar{M}| \bar{s}_1\bar{s}_2\bar{S})(0\bar{M}\bar{M}|\bar{\ell}\bar{S}J)  \nonumber\\
& &\times~{\cal A}^{(I)}(p_1,p_2,p_3,p_4;\epsilon_1,\epsilon_2,\epsilon_3^{\ast},\epsilon_4^{\ast})\ ,
\end{eqnarray}
where $M=\sigma_1+ \sigma_2$ and $\bar{M}=\bar{\sigma}_1+\bar{\sigma}_2$ with $\sigma_i$ is the third component of spin $s_i$ in the center-of-mass frame, $m$ is the third component of orbital angular momentum $\ell$. The Clebsch-Gordan coefficient $(a_1a_2A|b_1b_2B)$ is the composition for $b_1+b_2=B$, with $a_i$ and $A$ referring to the third components of the $b_i$ and $B$. The expressions of three-momentum are
\begin{eqnarray}
 \mathbf{p}_1=|\mathbf{p}|\hat{\mathbf{z}} , \quad
 \mathbf{p}_2=-|\mathbf{p}|\hat{\mathbf{z}} , \quad
 \mathbf{p}_3=\mathbf{p}'' , \quad
 \mathbf{p}_4=-\mathbf{p}'' .
 \end{eqnarray}
The left-hand cut will appear when the exchanging particles in the crossed channels as indicated in Eq.~\eqref{ampt} become on-shell in the partial-wave integration of Eq.~\eqref{eq:pwd}. For a $t$-channel exchanging term, its typical partial-wave integral gives 
\begin{widetext}
\begin{eqnarray}
&&\frac{1}{2} \int_{-1}^{+1} d\cos  \theta  \frac{1}{t-M_{ex}^2+i\epsilon} 
 = -\frac{s}{\sqrt{\lambda(s,M_1^2,M_2^2)\lambda(s,M_3^2,M_4^2)}}   \nonumber\\
&&\quad \quad \quad \times  \log \left[ \frac{M_1^2+M_3^2-(s+M_1^2-M_2^2)(s+M_3^2-M_4^2)/(2s)-
\sqrt{\lambda(s,M_1^2,M_2^2)\lambda(s,M_3^2,M_4^2)}/(2s)-M_{ex}^2+i\epsilon}{M_1^2+M_3^2-(s+M_1^2-M_2^2)(s+M_3^2-M_4^2)/(2s)+
\sqrt{\lambda(s,M_1^2,M_2^2)\lambda(s,M_3^2,M_4^2)}/(2s)-M_{ex}^2+i\epsilon}\right],
\label{eq:log}
 \end{eqnarray}
\end{widetext} 
where the K\"allen function $\lambda(x,y,z)=x^2+y^2+z^2-2xy-2yz-2xz$. The branch point of the left-hand cut can be easily obtained from Eq.~\eqref{eq:log} by requiring the argument of log vanishes.

With the partial-wave projected potentials $V^{(IJ)}$, one can obtain the unitarized amplitude $T$ by on-shell factorization
\begin{equation}
T^{(IJ)}(s)=\left[1-V^{(IJ)}(s)G(s)\right]^{-1}V^{(IJ)}(s) \ ,
\label{eq:BS}
\end{equation}
where the two-meson loop function $G(s)$ is
\begin{equation}
G(s)=i\int\frac{d^4q}{(2\pi)^4}\frac{1}{q^2-m_1^2+i\epsilon}\frac{1}{(q-P)^2-m_2^2+i\epsilon}\ ,
\label{eq:loopex}
\end{equation}
with $m_1$ and $m_2$ the masses of the two mesons involved in the loop, and $P$ is the total four-momentum of the meson-meson system. The two-point loop function $G(s)$ is logarithmically divergent and can be calculated with a once-subtracted dispersion relation whose explicit expression is~\cite{Oller:1998zr,Oller:2000fj}
\begin{eqnarray}
G(s)&=&\frac{1}{16\pi^2}\left[\alpha(\mu)+\log\frac{m_1^2}{\mu^2}+\frac{m_2^2-m_1^2+s}{2s}\log\frac{m_2^2}{m_1^2}\right. \nonumber\\
&&+\frac{q_\text{cm}}{\sqrt{s}}\left(\log\frac{s-m_2^2+m_1^2+2q_\text{cm}\sqrt{s}}{-s+m_2^2-m_1^2+2q_\text{cm}\sqrt{s}}\right. \nonumber \\
&&+\left. \left. \log\frac{s+m_2^2-m_1^2+2q_\text{cm}\sqrt{s}}{-s-m_2^2+m_1^2+2q_\text{cm} \sqrt{s}}\right)\right] ,
\label{eq:loopexdm}
\end{eqnarray}
where $\alpha(\mu)$ is the subtraction constant  depending on the renormalization scale $\mu$,  $q_\text{cm}$ is the magnitude of the three-momentum of the meson in the center of mass frame 
\begin{equation}
q_\text{cm}=\frac{\sqrt{\left[s-(m_1+m_2)^2\right]\left[s-(m_1-m_2)^2\right]}}{2\sqrt{s}} .
\end{equation}
Alternatively, one can also calculate the loop function by using a hard cutoff~\cite{Oller:1997ti}
 \begin{equation}
G_c(s)=\int_0^{q_{\mathrm{max}}} \frac{q^2 dq}{(2\pi)^2} \frac{\omega_1+\omega_2}{\omega_1\omega_2 [s-(\omega_1+\omega_2)^2+i\epsilon]   } \ ,\label{eq:loopexcut}
\end{equation}
where $q_{\mathrm{max}}$ is the cutoff of the three-momentum, $\omega_i=\sqrt{\vec{q}\,^2+m_i^2}$. The natural value of the cutoff is the scope of the low-energy theorem, or the scale of chiral symmetry breaking, e.g., $q_\text{max}\simeq \Lambda_\chi \simeq M_V$. Then we can get a natural value for the subtraction constant $\alpha$ in Eq.~\eqref{eq:loopexdm} by matching $G(s)$ and $G_c(s)$ at the threshold \cite{Gulmez:2016scm}. By setting $\mu=1500~\text{MeV}$ as in Ref.~\cite{Molina:2022jcd}\footnote{Note that the choice of $\mu$ in Eq.~\eqref{eq:loopexdm} is arbitrary as its variation can be fully absorbed by the redefinition of $\alpha$.}, the corresponding range of the value of $\alpha$ is determined to be $(-0.75,-1.65)$ by matching the $G_c(s)$ with $500~\text{MeV}<q_\text{max}<1300~\text{MeV}$ for the $D^*K^*$ loop. Unitarity leads to a cut along the real axis above the corresponding two-body threshold for each loop function, which divides the energy plane into two RSs. The expressions in Eqs.~\eqref{eq:loopexdm} and \eqref{eq:loopexcut} are for the first (physical) RS, denoted by $G^I(s)$, and the analytic continuation to the second RS can be obtained via~\cite{Oller:1997ti}
\begin{equation}
\label{eq:loopex2}
G^{II}(s)=G^{I}(s)+i\frac{q_\text{cm}}{4\pi\sqrt{s}}.
\end{equation}
For a $n$-channel system, there are $2^n$ RSs in total. Various sheets can be accessed by different choices of the loop functions $G^{I/II}(s)$ for each channel. In particular, for the $D_s^*\rho$-$D^*K^*$ system, there exist 4 RSs, which are labeled as $\{ 1,1\}$, $\{2,1\}$, $\{2,2\}$, and $\{1,2\}$. The first one is the physical sheet, while the last three are unphysical ones. The RS=$\{2,1\}$ connects to the physical one through the interval between the $D_s^*\rho$ and $D^*K^*$ thresholds, and the RS=$\{2,2\}$ is connected to the physical region above the $D^*K^*$ threshold along the real axis. Although the RS=$\{1,2\}$ is not directly connected to the physical region, a pole in it can still leave an impact on the physical observables due to the proximity of the $D^*_s\rho$ and $D^*K^*$ thresholds. 

The poles of the unitarizied amplitude could be identified as possible states. Poles located on the real axis below the lowest threshold in the physical RS correspond to possible bound states, and those on the unphysical RSs correspond to resonances. In particular, the poles on the real axis in unphysical RSs below the lowest threshold are called virtual states. A virtual state (as well as a resonance) does not correspond to a spatially localized state. However, such a pole can leave a significant imprint on the line shapes at the threshold if located near the threshold. The poles of the $T$-matrix correspond to the zeros of the determinant
\begin{equation}
\mathrm{Det}(s)=\mathrm{det} \left[1-V (s)G(s)\right].
\label{eq:Det}
\end{equation}
In addition, we can define an effective coupling of the channel $i$ ($j$) to a given state at the pole $s_0$ by the residues of the transition amplitude $T_{ij}$ via
\bea
g_ig_j = \lim_{s\to s_0}\left(s-s_0\right)T_{ij}(s).
\eea


\renewcommand\arraystretch{1.5}
 \begin{table*}[thp]
\caption{The pole positions and effective couplings evaluated for $I = 1$, $J = 0$ on different RSs with $\mu=1500~\mathrm{MeV}$. The threshold of $D_s^*\rho$ is 2887~MeV.}
\label{tab:Tcs2900}	
\begin{tabular}{p{2.5cm}<\centering  p{2.5cm}<\centering  p{3.5cm}<\centering   p{3.5cm}<\centering  p{3.5cm}<\centering}
\toprule[1pt]
 RS & $\alpha$ & $\sqrt{s}_{\mathrm{pole}}$ [MeV] & $|g_{D^{\ast} K^{\ast}}|$ [MeV]  & $|g_{D_s^{\ast} \rho}|$ [MeV] \\
\midrule[1pt]
 \{1,1\}	& -1.65$\sim$-1.60 & 2885$\sim$2887  & 5531$\sim$2198  & 5379$\sim$2082 \\
\{2,1\}	& -1.60$\sim$-1.55 & 2887$\sim$2885  & 1755$\sim$8202  & 1650$\sim$7348 \\
\{1,2\} & -1.39$\sim$-1.35 & 2885$\sim$2887  & 6587$\sim$1625  & 7886$\sim$1865 \\
 \{2,2\}	& -1.35$\sim$-1.28 & 2887$\sim$2885  & 1415$\sim$4202  & 1613$\sim$4672 \\
\bottomrule[1pt]   	
\end{tabular}
\end{table*}

  \begin{figure*}[htp]
  \centering
   \includegraphics[scale=0.9]{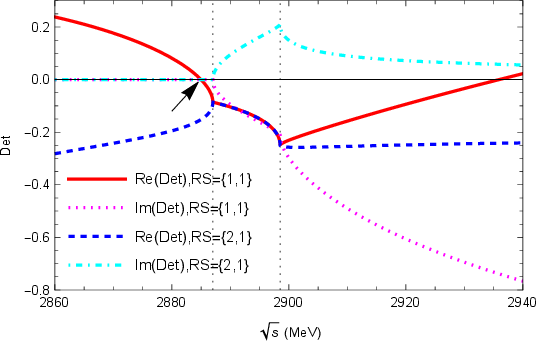}\put(-20,135){(a)}
   \includegraphics[scale=0.9]{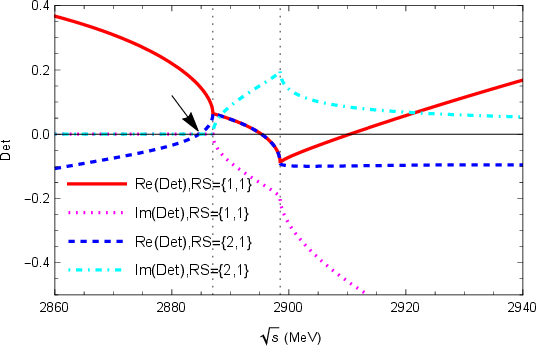}\put(-20,135){(b)}
   \caption{The determinant defined by Eq.~\eqref{eq:Det} for $I=1$, $J=0$ with $\alpha=-1.65$ (left) and $\alpha=-1.55$ (right). The red solid line: the real part of the determinant on the Riemann sheet $\mathrm{RS}=\{1,1\}$, and the magenta dotted line: the imaginary part of the determinant on $\mathrm{RS}=\{1,1\}$, the blue dashed line: real part of determinant on $\mathrm{RS}=\{2,1\}$,  the cyan dot-dashed line: imaginary part of determinant on $\mathrm{RS}=\{2,1\}$, the lower black dotted line: the $D^{\ast}_s \rho$ threshold, and the upper black dotted line: the $D^{\ast} K^{\ast}$ threshold. The arrows refer to the position of the bound state (left) and virtual state (right).}  
   \label{fig:det1121}
  \end{figure*}

  \begin{figure*}[htp]
  \centering
   \includegraphics[scale=0.9]{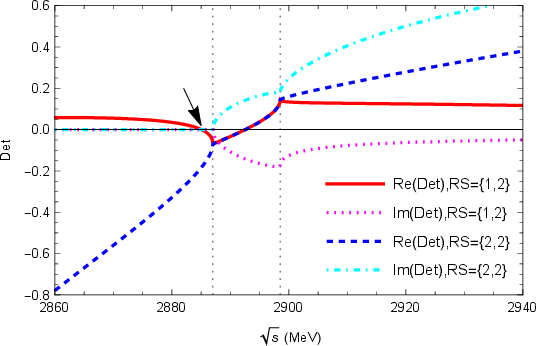}\put(-205,135){(a)}
   \includegraphics[scale=0.9]{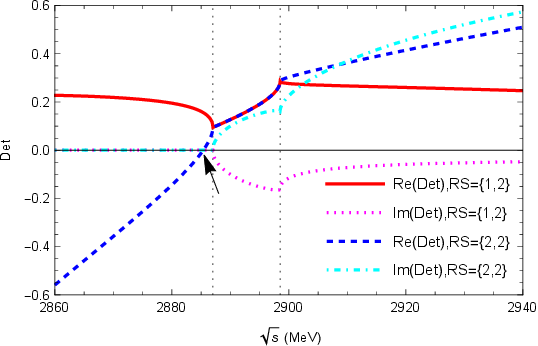}\put(-205,135){(b)}
   \caption{The same as in Fig.~\ref{fig:det1121}, except that the determinant with $\alpha=-1.39$ (left) and $\alpha=-1.28$ (right) on $\mathrm{RS}=\{1,2\}$ and $\mathrm{RS}=\{2,2\}$. The arrows refer to the position of the virtual states.}  
   \label{fig:det1222}
  \end{figure*}

\section{NUMERICAL RESULTS AND DISCUSSIONS}
\label{sec:RESULTS}

It is noticed that either a bound state or a virtual state can be formed in the aforementioned range of $\alpha$, i.e. $-1.65<\alpha<-0.75$, for the $J=0$ isovector with both $g=4.17$ and $4.60$. It implies that the $\tcs$ is consistent with a $D_s^*\rho$-$D^*K^*$ bound state/virtual state. In particular, for $\alpha=-1.60$ and $g=4.17$, we find a bound state using the relativistic potentials in Eq.~\eqref{eq:totalamp} with the binding energy $E_B=0.03$ MeV with respect to the $D_s^*\rho$ threshold. However, by employing the nonrelativistic potentials in Ref.~\cite{Molina:2010tx}, a virtual state very close to the $D_s^*\rho$ is found in the RS=$\{2,1\}$ with masses 1.86 MeV below the $D_s^*\rho$ threshold. This virtual state is located in the vicinity of the $D_s^*\rho$ threshold thus produces a cusp in the threshold. By employing $g=4.60$, the virtual state resulting from the nonrelativistic potentials turns out to be a bound state in the physical RS with a binding energy of 13.70 MeV. We stress the distinction between the two scenarios of the relativistic and the nonrelativistic potentials beyond the accuracy of the frameworks used. However, it is quite certain that the $D_s^*\rho$-$D^*K^*$ interaction is attractive, at least near the thresholds, which hints at the existence of a pole near the thresholds. 

In the present work, we assume that the $\tcs$ corresponds to the $D_s^*\rho$-$D^*K^*$ bound/virtual state discussed above. Based on that we determine the unknown constant $\alpha$ from the Breit-Wigner mass of the $\tcs$. In what follows, we employ $g=4.17$ as in Ref.~\cite{Molina:2010tx}. The results for the $g=4.60$ is similar since the effect from the change of the $g$ is largely compensated by the adjustment of the subtraction constant $\alpha$. In addition, we employ the relativistic potentials which have left-hand cuts originating from the vector-exchanging diagrams. In Figs.~\ref{fig:V0} and ~\ref{fig:V1}, we present the potentials with $I=0$ and $I=1$, respectively. From these figures, one can find that our results and the ones of Ref.~\cite{Molina:2010tx} are similar typically near the $D^* K^*$ threshold, however for lower values of $\sqrt{s}$, they depart quickly due to relativistic corrections and the onset of the left-hand cuts. As shown in Figs.~\ref{fig:V0}-(a1), (a2), and (a3), the peculiar structures around the left-hand cuts of the $D^{\ast}K^{\ast} \to D^{\ast}K^{\ast}$ channel with $I=0$ are derived from the effect of exchanging $\rho$ and $\omega$ particles in the $t$ channel amplitude, which also appear in Figs.~\ref{fig:V1}-(f1), (f2),  and (f3). The left-hand cuts of Figs.~\ref{fig:V0}-(a), (b), (d), (e) and \ref{fig:V1}-(f), (h) appear at 2772~MeV, 2722~MeV, 2836~MeV, 2558~MeV, 2772~MeV, and 2718~MeV, respectively. Notice that the presence of the left-hand cuts invalidates the on-shell factorization employed in Eq.~\eqref{eq:BS}. As a result, we restrict ourselves to the energy region above the left-hand cuts. It is reasonable since the corrections from the relativistic kinematics and the higher-order effective Lagrangian could be significant for the regions far below the threshold. To be concrete, we only consider the poles above $\sqrt{s}=2840$~MeV for $I=0$ and above $\sqrt{s}=2780$~MeV for $I=1$. 

\begin{figure*}[htb]
\centering
\includegraphics[width=1\textwidth]{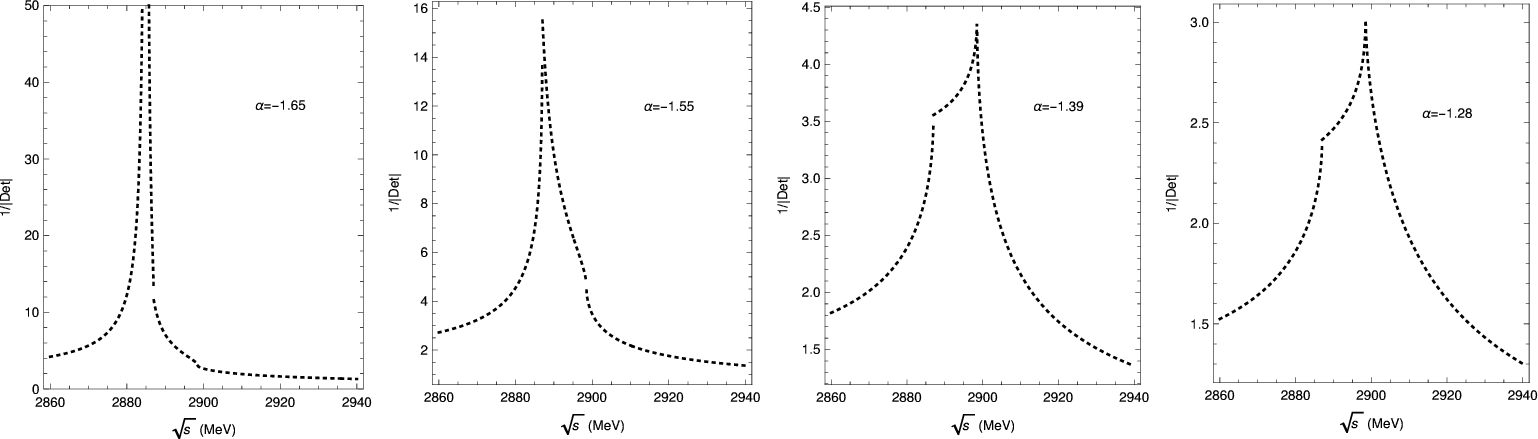}
\caption{$1/|\text{Det}|$ in the physical RS with four different values of $\alpha$ which produces a pole at 2885~MeV in RS=$\{1,1\}$, $\{2,1\}$, $\{1,2\}$ and $\{2,2\}$, respectively. }  
\label{fig:dets}
\end{figure*}

Starting from $\alpha=-1.65$, corresponding to the $G_c(s)$ with $q_\text{max}=1300$~MeV, one finds a pole at 2885 MeV for the $(I,J)=(1,0)$ sector in the physical sheet, as shown in the left panel of Fig.~\ref{fig:det1121}, which is accidentally on the top of the edge of the 1$\sigma$ uncertainty band of the $T_{c\bar{s}0}(2900)$ mass $m_{T_{c\bar{s}0}(2900)}=(2908\pm11\pm20)$~MeV. By increasing the $\alpha$, the pole corresponding to a bound state moves towards the $D_s^*\rho$ threshold and hits the threshold at $\alpha=-1.60$. Then it turns into a virtual state in RS=$\{2,1\}$ and the pole position moves away from the threshold towards the left-hand cuts with increasing $\alpha$ and arrives at 2885~MeV again with $\alpha=-1.55$, see e.g. the right panel of Fig.~\ref{fig:det1121}. Keep increasing $\alpha$, a pole in the RS=$\{1,2\}$ can be found in the real axis below the $D_s^*\rho$ threshold, and it moves towards the threshold. In particular, at $\alpha=-1.39$, the pole in RS=$\{1,2\}$ is located at 2885~MeV. Meanwhile, the pole in RS=$\{2,1\}$ moves to 2813~MeV. Increasing $\alpha$ to $-1.35$, the pole in RS=$\{1,2\}$ hits the $D_s^*\rho$ threshold and turns into a pole in RS=$\{2,2\}$, where the pole position moves from the threshold to the left-hand cut and arrives at 2885~MeV for $\alpha=-1.28$, see e.g. in Fig.~\ref{fig:det1222}. As mentioned above, a pole in the vicinity of the threshold in an unphysical sheet could also leave an impact on the physical observables. To see that, $1/|\text{Det}|$ evaluated in the physical RS are shown in Fig.~\ref{fig:dets} with four different values of $\alpha$, which produce a pole at 2885~MeV in RS=$\{1,1\}$, $\{2,1\}$, $\{1,2\}$ and $\{2,2\}$, respectively. Therefore, by identifying the $\tcs$ as a $(I,J)=(1,0)$ $D_s^*\rho$-$D^*K^*$ bound/virtual state, we obtain a range of the parameter $\alpha$ from the Breit-Wigner mass of the $\tcs$ under the uncertainty given by Eq.~\eqref{respara}, i.e., $-1.65<\alpha<-1.55$ and $-1.39<\alpha<-1.28$. The corresponding pole positions and the effective couplings are collected in Table~\ref{tab:Tcs2900}. Two disconnected intervals of $\alpha$ are caused by the mass splitting of $D_s^*\rho$ and $D^*K^*$. If one approaches the SU(3) symmetry and decreases the mass splitting of the two channels, the difference between the RS=$\{1,1\}$ and $\{1,2\}$ diminishes. And under the exact SU(3) symmetry, there are only two RSs surviving, i.e., $\{1,1\}$ and $\{2,2\}$, in which case, the two intervals of $\alpha$ coincide.

With the parameter $\alpha$ in hand, we are equipped to investigate the sectors $I=1$, $J=1,~2$ and $I=0$, $J=0$, 1, 2. The poles found in these sectors are collected in Table~\ref{tab:rangeS}. As for $I=1$ and $J=1$, a pole is found located at 2886~MeV on RS=$\{1,1\}$ with $\alpha=-1.65$. Its pole mass increases to the $D^{\ast}_s \rho$ threshold and then decreases once it arrives at the threshold with the $\alpha$ variation from $-1.65$ to $-1.55$. For $-1.39<\alpha<-1.36$, it should be noted that two virtual states are found in RS=$\{1,2\}$. However, only the one with higher mass which is closer to the physical region can leave significant imprints on the observable, and thus are kept in Table~\ref{tab:rangeS}. Similarly, we do not show the poles which are far from the physical region and do not impact the line shapes. For the sector of $I=1$ and $J=2$, we find poles in the physical RS with the pole mass $2780\sim2806$~MeV for the $\alpha$ in the interval ($-1.31,-1.28$). The mass is consistent with that predicted in Ref.~\cite{Molina:2010tx} with the pole position $2786$~MeV in the sector of $C=1$, $S=1$, $I=1$ and $J=2$. We should mention, however, that we do not predict a bound state with mass $2780$ to $2806$ MeV, which is only found with a certain range of determined $\alpha$ from the $\tcs$. For the rest of the $\alpha$ values, we do not find a pole above 2780 MeV. Similarly, we do not find poles above the left-hand cuts for sectors $I=0$, $J=0$, 1 and 2. 
For the region below the left-hand cuts, it beyonds the capability of the current effective Lagrangian and the on-shell factorization as mentioned above \cite{Gulmez:2016scm,Du:2018gyn}.

So far we have neglected the widths of the vector mesons and the inelastic channels, which will generate widths for those bound states and virtual states mentioned above and turns them into resonances. The significant decay widths of $\rho\to\pi\pi$, $K^*\to K\pi$ imply that the contributions from the $D_s^*\pi\pi$, $D^*K\pi$ three-body (and even $DK\pi\pi$ four-body) intermediate states to the widths of the generated states should be the order of the width of $\rho$/$K^*$. In addition, the pseudoscalar intermediate states, e.g. the $D_s\pi$ and $DK$, contribute to the widths as well, corresponding to the decays into these two mesons. In order to take such contributions into account, one has to introduce model-dependent form factors. While the width of the generated resonance is sensitive to the form factors, the real part of the pole position is merely affected \cite{Molina:2008jw,Gulmez:2016scm}. In the present work, we focus on the origin of the poles and their masses, and do not consider the convolution of loop functions accounting for the widths of $\rho$ and $K^*$ in their propagators and box diagrams assessing the $D_s\pi$ and $DK$ inelastic contributions~\cite{Liu:2023hrz,Garcia-Recio:2013uva}.



\renewcommand\arraystretch{1.5}
 \begin{table*}[thb]
\caption{The pole positions evaluated in the sectors of $I = 1, J = 1$ and $I = 1, J = 2$ on the different RSs with $-1.65 < \alpha < -1.55$ and $-1.39 < \alpha < -1.28$. The ``-" indicates that no pole is found. In the present work, we only consider the energy region safe from the left-hand cut, i.e. $\sqrt{s}>2780$ MeV.}
\label{tab:rangeS}	
\begin{tabular}{p{3cm}<\centering  p{3cm}<\centering  p{3cm}<\centering p{0.2cm}<\centering  p{3cm}<\centering  p{3cm}<\centering}
\toprule[1pt]
 \multirow{2}{*}{RS} & \multicolumn{2}{c }{I=1, J=1} & &\multicolumn{2}{c}{I=1, J=2} \\
 \cmidrule[1pt]{2-3}  \cmidrule[1pt]{5-6}
 \, & $\alpha$ & $\sqrt{s}_{\mathrm{pole}}$ [MeV] && $\alpha$ & $\sqrt{s}_{\mathrm{pole}}$ [MeV]\\
\midrule[1pt]
 \{1,1\} & $-1.65 \sim -1.61$ & $2886 \sim 2887$ & &$-1.31 \sim -1.28$ & $2780 \sim 2806$\\
\{2,1\} & $-1.61 \sim -1.55$ & $2887 \sim 2883$ &&- &-\\
  \{1,2\} & $-1.39 \sim -1.36$ & $2886 \sim 2887$&&- &-\\
\{2,2\} & $-1.36 \sim -1.28$ & $2887 \sim 2885$ &&- &-\\
\bottomrule[1pt]   	
\end{tabular}
\end{table*}

\section{SUMMARY}
\label{sec:SUMMARY}

The recently observed spin-party $J^P=0^+$ states $\tcsn$ and $\tcspp$ by the LHCb Collaboration in the $D_s^+ \pi^-$ mass distribution of the process $B^0 \to \bar{D}^0 D_s^+ \pi^-$ and the $D_s^+ \pi^+$ distribution of the $B^+ \to D^- D_s^+ \pi^+$, respectively, are in good agreement and belong to an isospin triplet \cite{LHCb:2022xob,LHCb:2022bkt}. By investigating the $D^*K^*$ coupled-channel system within the framework of the local hidden gauge approach extended to SU(4), we found that the $D_s^*\rho$-$D^*K^*$ system in the $(I,J)=(1,0)$ sector manifests a sizable attractive interaction which can form a bound or a virtual state within a reasonable parameter range. We have derived the scattering potentials including the relativistic corrections in a covariant formalism which develops left-hand cuts. The existence of the left-hand cuts invalidates the on-shell factorization in the vicinity of the left-hand cuts and below. Therefore we only focus on the region above the left-hand cuts. Notice that for the energy below the left-hand cuts, the contributions from the relativistic correction and the higher-order effective Lagrangian could be significant such that beyond the capability of the framework.

By assuming the $\tcs$ as a $D_s^*\rho$-$D^*K^*$ bound/virtual state in the sector $(I,J)=(1,0)$ and reproducing the pole mass within the 1$\sigma$ uncertainty, we have determined the subtraction constant $\alpha$ to be in two intervals, i.e, $-1.65<\alpha<-1.55$ and $-1.39<\alpha<-1.28$, for the renormalization scale $\mu=1500$~MeV. With the subtraction constant, we have searched for the possible poles in the sectors of $I=1$, $J=1$, 2 and $I=0$, $J=0$, 1, 2. The sector $(I,J)=(1,1)$ has a comparable attraction with $(I,J)=(1,0)$ such that form a bound/virtual state in the range $(2883,2887)$ MeV. For the sector $(I,J)=(1,2)$, a stronger attraction interaction could generate a deeper bound state below 2806 MeV. For the isospin scalar sector, no pole can be found above the left-hand cuts where the framework works. The results in this work provide a good reference for the experimental studies of $(C,S)=(1,1)$ systems and help us to unravel the nature of the recently observed $\tcs$.

\section*{Acknowledgement}
This work is supported by the National Natural Science Foundation of China under Grant Nos .11775050, 12175037, and 12192263. This work is also supported by the Natural Science Foundation of Henan under Grand Nos. 222300420554 and 232300421140,  the Project of Youth Backbone Teachers of Colleges and Universities of Henan Province (2020GGJS017), and the Open Project of Guangxi Key Laboratory of Nuclear Physics and Nuclear Technology, No. NLK2021-08.


\end{document}